\documentclass[]{spie}
\usepackage[]{graphicx}

\title{Crystalline Whispering Gallery Mode Resonators: In Search of The Optimal Material}

\author{
V. S. Ilchenko, A. A. Savchenkov, A. B. Matsko, and L. Maleki
\skiplinehalf OEwaves Inc., 465 N. Halstead Street, Suite 140, Pasadena, California, 91107, USA}

\authorinfo{Send correspondence to A.B.Matsko:
Andrey.Matsko@oewaves.com}

\begin{document}

\maketitle

\begin{abstract}
Different applications of crystalline whispering gallery mode resonators call for different properties of the resonator host material. We report on our recent study of  resonators made out of sapphire, diamond, and quartz crystals and discuss possible applications of these resonators. In particular, we demonstrate Kerr frequency comb generation in sapphire microresonators.
\end{abstract}

\keywords{Diamond, Sapphire, Quartz, Whispering Gallery Modes, Crystalline Microresonator, Hyper-Parametric Oscillator, RF Photonic Oscillator, Self-Injection Locked Laser; \\ \\
{\small Reprinted from:
V. S. Ilchenko, A. A. Savchenkov, A. B. Matsko and L. Maleki, \\
 ``Crystalline whispering gallery mode resonators: in search of the optimal material,'' \\
 Proc. SPIE 8960, Laser Resonators, Microresonators, and Beam Control XVI, 896013 (March 4, 2014); \\ doi:10.1117$/$12.2044823; $http://dx.doi.org/10.1117/12.2044823$}
}

\section{Introduction}

An efficient technique of measurement of attenuation of transparent optical crystals involves fabrication of monolithic resonators out of the materials and measurement of their intrinsic quality ($Q-$) factors. The attenuation of light in the resonator mode, $\alpha$, is related to the quality factor of a circular monolithic resonator as $Q=2\pi n_0/(\lambda \alpha)$, where $\lambda$ is the wavelength of measurement, and $n_0$ is the refractive index of the material at this wavelength. The value of coefficient $\alpha$ is determined by both absorption and scattering in the bulk material and on the surface of the resonator. Hence, $\alpha$ gives a reasonable approximation of the bulk
absorption if other loss mechanisms are made negligible.

Whispering gallery mode (WGM) resonators are solid state cavities that confine light in small geometrical volumes for long periods of time \cite{ilchenko05jstqe}. These structures are promising for many applications in nonlinear optics, RF photonics, and optical technology \cite{ilchenko05jstqe} ranging from cavity QED \cite{aoki06n} to Kerr optical frequency comb generation \cite{delhaye07n,savchenkov08prl}. Finesse of WGM resonators can exceed the finesse of the best Fabry-Perot resonators and the linewidth of the WGMs can be as small as a few kHz at room temperature \cite{savchenkov07oe}.

Various properties of the resonators are required depending on particular applications. For example, planar on-chip integration of a microresonator calls for resonators with high enough index of refraction. The refractive index should exceed the refractive index of fused silica substrate and preferably approach refractive index of silicon nitride or silicon. Low index materials, like magnesium and calcium fluoride, are not suitable for this purpose. They hinders integration of Kerr frequency comb generators \cite{savchenkov08prl}. In this work we demonstrate, for the first time, generation of the Kerr frequency combs in sapphire resonators. Those resonators have excellent mechanical properties and represent a great potential for heterogeneous integration with photonic chips.

Ultra-stable cavities are required for laser stabilization \cite{liang10ol}. The ultimate stability of a monolithic microresonator is limited by fundamental thermodynamic fluctuations existing even if the resonator is kept at constant
temperature \cite{matsko07josab}. The thermal conductivity of the resonator host material is essential to reduce those fluctuations. In addition, resonators with high thermal conductivity are less sensitive to various kinds of thermal effects resulting from slow heat spread from the mode volume to the substrate \cite{gorodetsky92lp,carmon4oe,fomin05josab}. In this work we report on fabrication of a diamond resonator that has extremely high thermal conductivity \cite{ilchenko13ol}.

Tunability of WGM resonators is important in many applications. For instance, structures containing multiple interacting resonators require coincidence of the WGM frequencies belonging to different resonators \cite{astratov04apl,poon06ol,mookherjea07ol}. The fabrication of completely identical very high-Q resonators is not practical, so the postproduction trimming as well as fast real time tuning of the resonant frequencies is required. Some crystalline WGM resonators possessing electro-optic effect can be tuned electrically \cite{savchenkov03el}. However, the quality (Q) factors of those resonators are usually not very high and do not exceed one billion. Ultra-high Q WGM resonators are fabricated from amorphous or centro-symmetric crystals possessing some cubic nonlinearity and zero quadratic nonlinearity \cite{savchenkov07oe}. Such resonators can be tuned by either strain or heat, but cannot be efficiently electrically tuned. We demonstrate that crystalline quartz WGM resonators can bridge this gap. Quatrz resonators can simultaneously posses very high Q and be tuned electrically. We describe experimental demonstration of such resonators and discuss their properties and potential applications \cite{ilchenko08ol}.

In what follows we describe our results on studies of sapphire, diamond, and quartz whispering gallery mode resonators.

\section{Sapphire Whispering Gallery Mode Resonator}

\begin{figure}[htb]
\centerline{
\includegraphics[width=14.5cm]{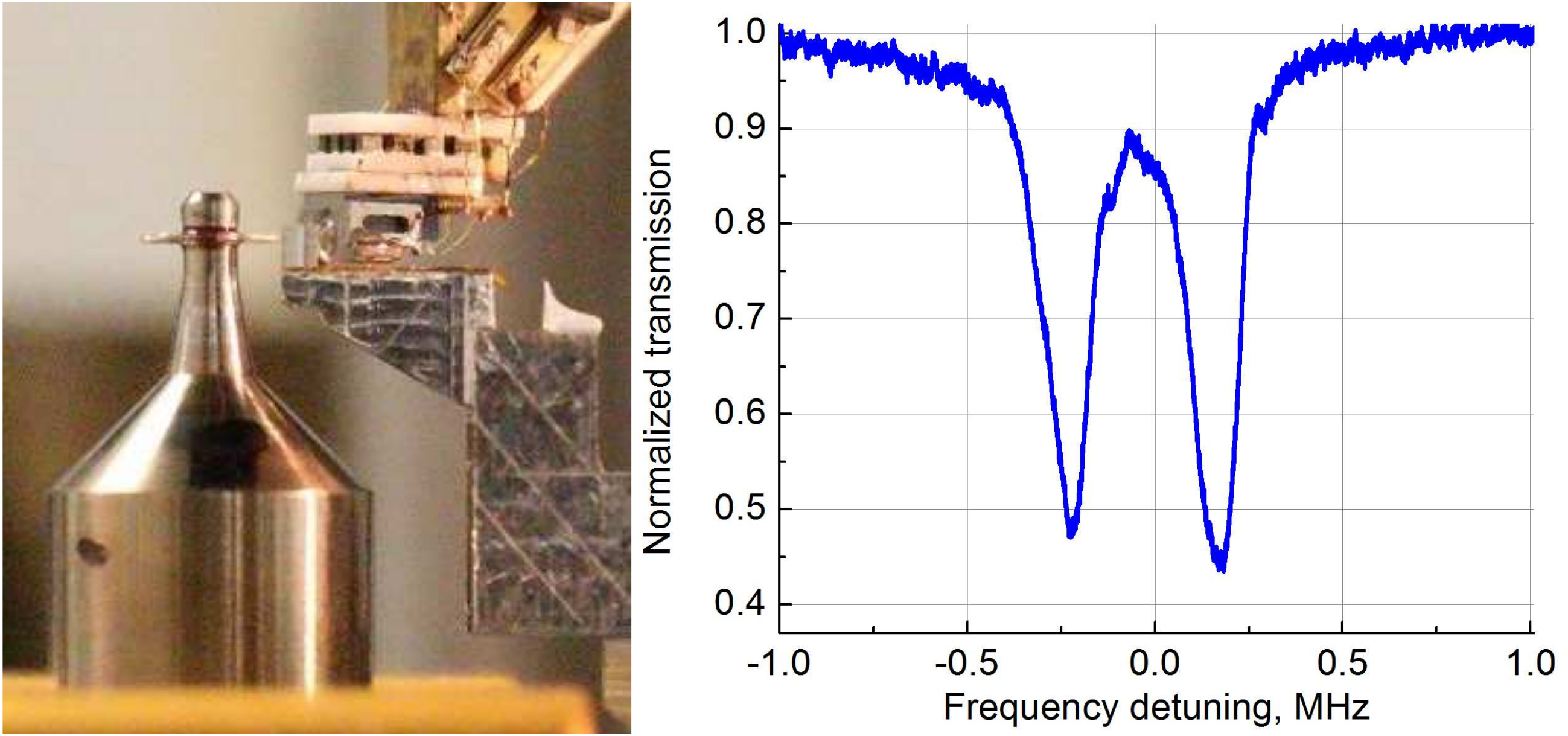}}
\caption{\label{fig1sapphire} A picture of the sapphire resonator on a manufacture pin inserted into the test setup (left panel), and example of the optical spectrum of the resonator (right panel). A distinct 420~kHz doublet due to Rayleigh scattering is clearly seen. The WGMs have bandwidths of 160~kHz for the polarization perpendicular to the resonator axis and 260~kHz for the polarization parallel to the resonator axis.}
\end{figure}

\begin{figure}[htb]
\centerline{
\includegraphics[width=14.5cm]{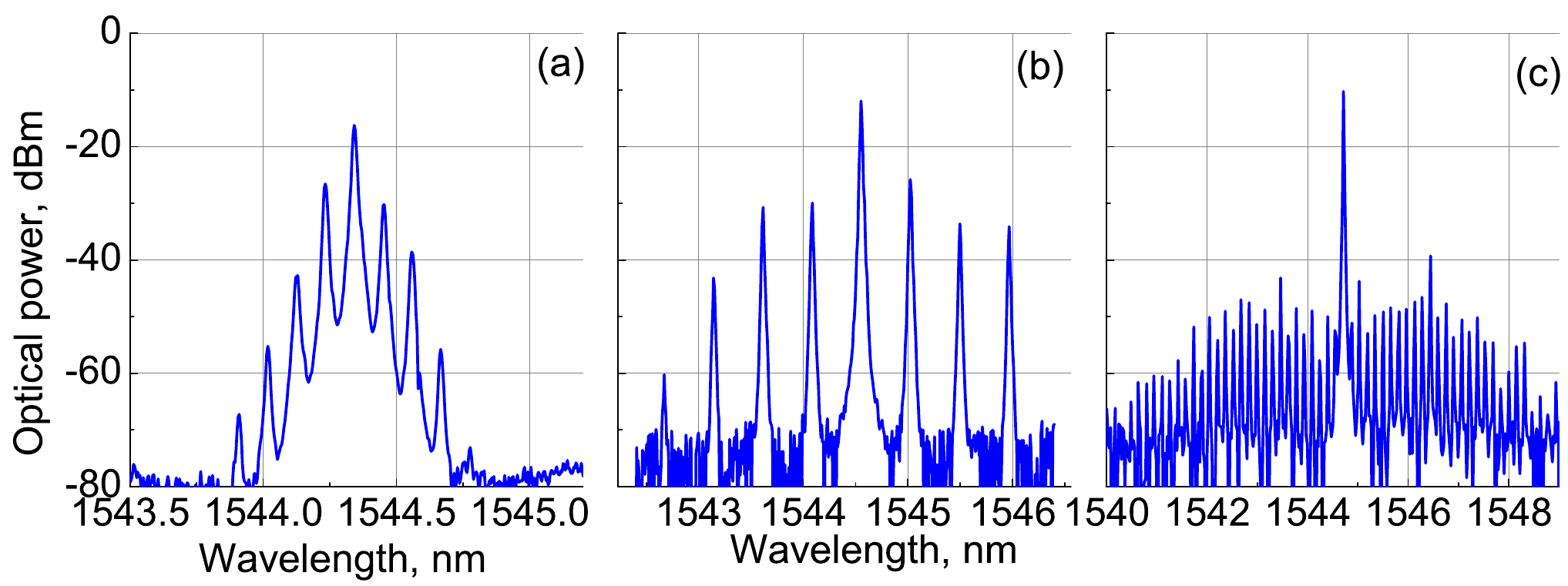}}
\caption{\label{fig2sapphire} Examples of optical spectra of Kerr frequency combs generated in the sapphire resonator.}
\end{figure}

We have fabricated sapphire whispering gallery mode resonators (Fig.~\ref{fig1sapphire}) out of sapphire preforms and shown that pumping the resonators with continuous wave light results in generation of Kerr frequency combs (Fig.~\ref{fig2sapphire}). The resonators have coupled Q-factor exceeding $10^9$ at 1550~nm wavelength. Strong stimulated Rayleigh scattering results in splitting the WGM spectra into doublets, shown at left panel of (Fig.~\ref{fig1sapphire}). Since sapphire has high enough index of refraction (1.74705 ordinary and 1.73924 extraordinary at 1500~nm) and is easily machinable, it is well suited for planar integration. Mid-IR transparency of the material makes it promising for generation of frequency combs in this spectral region.


\section{Diamond Whispering Gallery Mode Resonator}

While it is well known that diamond has an extremely broad optical transparency window, there is little consistent experimental data on its attenuation within the window. It is relatively easy to measure transparency of fused silica  because of kilometers of optical fibers fabricated from the material, however this approach is not applicable to the diamond by obvious reasons.

Methods of fabrication of high quality diamond crystals \cite{spearbook} has resulted in the material being used in a range of novel applications from optic to quantum information
\cite{tallaire06drm,awschalom07sa,balmer09jpcm,balasubramanian09nm,friel10spie,aharonovich11np}. The material is attractive because of its exceptional physical and chemical properties, including high thermal conductivity, mechanical hardness, and a wide optical transparency window from vacuum ultraviolet through the infrared.

The best Q-factor of a monolithic diamond resonator measured at 1,550~nm, up to our knowledge, is
2.5$\times10^5$ \cite{hausmann13nl}, which corresponds to linear attenuation coefficient $\alpha \simeq 0.4$~cm$^{-1}$. The measurements were performed with a single crystal diamond race-track resonator built on top of a
silicon dioxide/silicon substrate. Observations have included the splitting of relatively low Q
resonances due to stimulated Rayleigh scattering, hence we can conclude that the experimentally
measured Q values are not limited by the bulk loss of the material. The scattering usually results
from roughness of the surface of open microresonators. Mechanical polishing allows reaching
outstanding surface quality that already lead to demonstration of the highest finesse monolithic
resonators (${\cal F}>10^7$, \cite{savchenkov07oe}). We used this technique to fabricate a diamond
resonator \cite{ilchenko13ol}. The measured quality factor of the resonator was $Q=2.4 \times 10^7$, which is two orders of magnitude larger compared with the best previous measurement. We found that in our case the
measured Q-factor is limited by the material loss approaching $\alpha=4\times10^{-3}$~cm$^{-1}$ and
not by  surface scattering.

The single-crystal diamond samples used in this experiment were synthesized with a microwave
plasma-assisted chemical vapor deposition reactor operating at a frequency of 896~MHz. The sample was
produced from homo-epitaxial layer grown on a $\langle 1,0,0 \rangle$-orientated diamond surface that
had been prepared using high quality polishing techniques to minimize Ra ($<$20 nm) and therefore
reduce the nucleation of dislocations in the epitaxial layer.

The diamond that had approximately 20~ppb of nitrogen impurities in the solid, which results in extremely low levels of absorption across the whole transparency spectrum \cite{friel10spie}. This kind of a diamond was utilize to build the WGM resonator. Diamond parts for quantum computing applications have been produced in thin layers with nitrogen impurities below 1~ppb \cite{balasubramanian09nm}, and it can be foreseen that using the techniques exploited in the synthesis of these layers that in the future even lower absorption bulk diamond can be produced.

\begin{figure}
\centering\includegraphics[width=14.5cm]{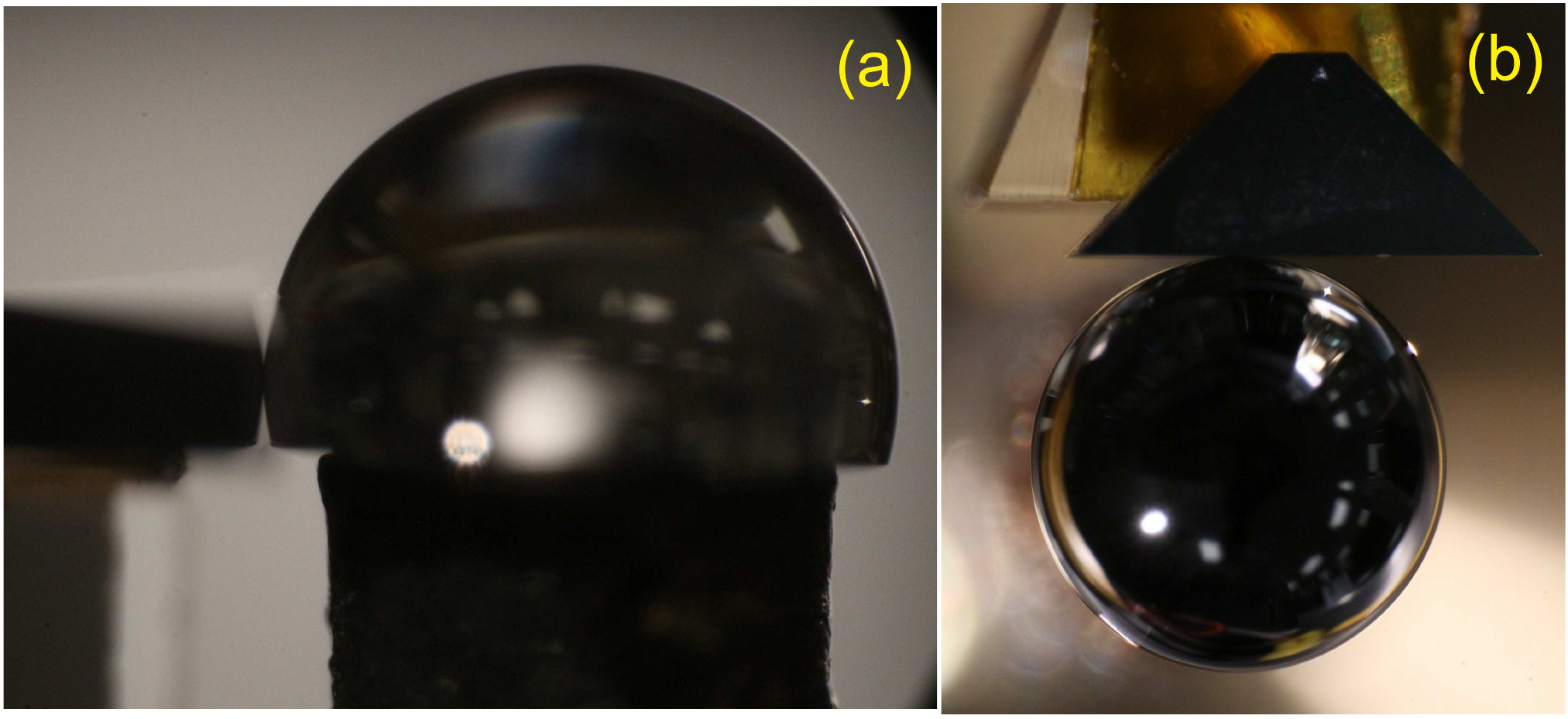}
\caption{\label{fig1diamond} Pictures of the fabricated diamond whispering gallery mode resonator: (a) side view, (b) top view. A silicon prism is used to couple light in and out of the resonator. }
\end{figure}

The fabricated resonator is shown in Fig.~(\ref{fig1diamond}). It is shaped as a  sphere with
the diameter of $2a=2$~mm,truncated  about 130 micron below large diameter ("equatorial plane") and
suspended on a metallic rod during fabrication and measurement. Refractive index of diamond is
$n_0=2.386$ at 1,550~nm.

\begin{figure}
\centering\includegraphics[width=14.5cm]{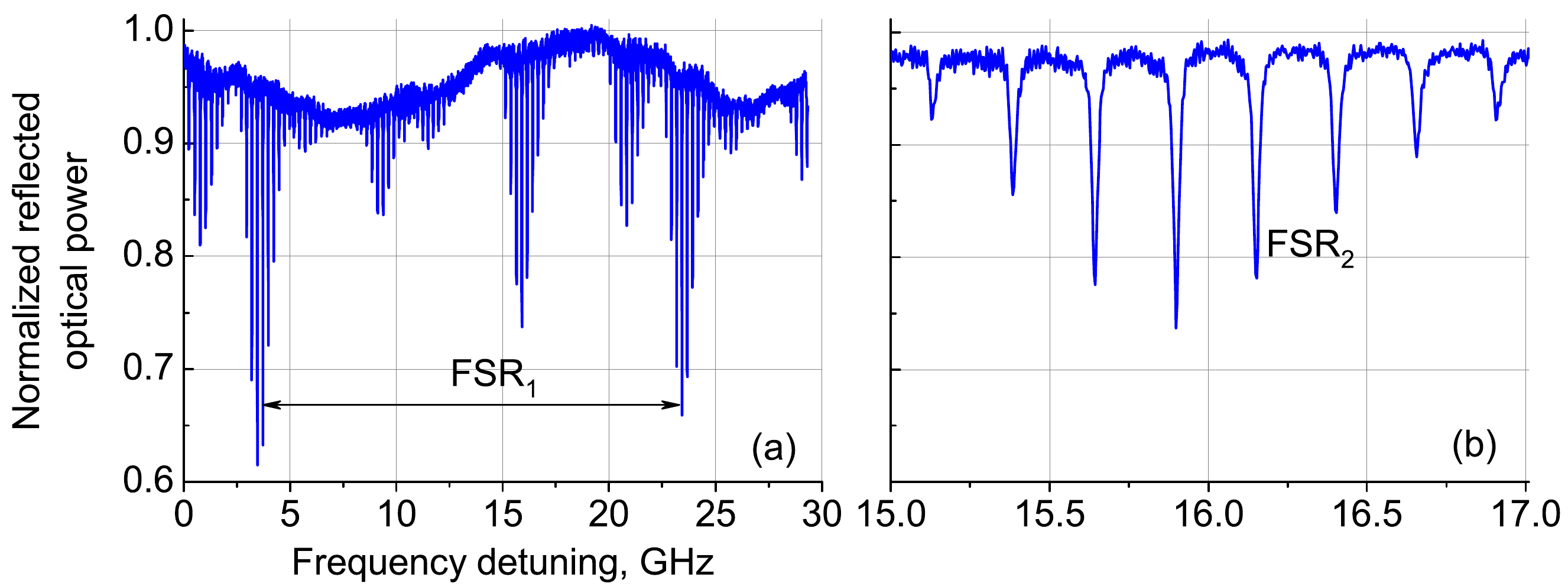}
\caption{\label{fig2diamond} Spectrum of the resonator modes polarized
perpendicular to the resonator symmetry axis. Two free spectral ranges
(FSRs)are observed. FSR$_1$ corresponds to azimuthal quantization of the
modes, FSR$_2$ corresponds to altitudinal quantization of the modes.}
\end{figure}

Measurement of the Q-factor was done by means of the prism coupling method.  We used a silicon prism
(refractive index $n_p=3.48$) to couple light in and out of the resonator modes. The ultimate
coupling efficiency achieved in the experiment exceeds 70\%. A tunable fiber laser was utilized as a
pumping light source. The linewidth of the emitted light ($<4$~kHz) was small enough to perform
spectroscopy of the resonator modes. A typical measured spectrum is shown in Fig.~\ref{fig2diamond}. The
intrinsic Q-factor was inferred from the measurement of the bandwidth of the observed resonances in
the undercoupled regime.

The modes are characterized with unloaded full width at the half maximum of $8-10$~MHz. A general
criterion to distinguish surface and bulk attenuation in the open optical resonators is the
equivalence of the unloaded quality factors of all the resonator modes independently on their
polarization. The modes polarized along the symmetry axis have slightly worse Q-factor (less than
10\%), which indicates that the surface of the resonator is either slightly damaged or contaminated.
Visual inspection has indicated presence of the minor residual scratches that can explain the slight
difference in the Q-factors because of different impact of surface scattering on Q of WGMs with
different polarization.

Nearly the same Q-factor was measured at 1319~nm wavelength using diode pumped Nd:YAG solid state
laser. The laser light had 100~kHz linewidth. It was coupled to the resonator using the same coupling
prism. This measurement shows that the attenuation is rather broadband.

The spectrum of the resonator has distinct azimuthal and latitudinal spectral features, which indicates
that the resonator has a slightly spheroidal shape. The azimuthal free spectral range is 20~GHz. The
latitudinal spectral range depends on the polarization of the modes. It is $263$~MHz for the modes
polarized collinearly with the symmetry axis of the resonator, and $249$~MHz for the modes polarized
perpendicularly to the axis. We were able to optimize the coupling with the higher order modes by
changing the launch angle for the beam of the pumping light impinging on the resonator surface.

A spheroidal resonator has two nearly equidistant mode sequences characterized with frequency
intervals
\begin{equation}
FSR_1=\frac{c}{2\pi n_0 a}
\end{equation}
and
\begin{equation}
FSR_2=FSR_1 \frac{a-b}{b}
\end{equation}
$a$ and $b$ are the semi-axes of the spheroid
\cite{sumetsky04ol,louyer05pra,gorodetskyo6jstqe,pollinger09prl}. Using result of our measurement we
conclude that non-sphericity of the resonator is approximately 1\% ($(a-b)/b \simeq 0.014$).

Thermal nonlinearity is an important factor that limits the optical power range in applications of
high-Q resonators \cite{gorodetsky92lp,carmon4oe,fomin05josab}. The easily observed manifestation of
thermal nonlinearity is the visible extension (or compression) of WGM resonances observed during
sweeping the probe laser: because of the thermal nonlinearity, the trace of the resonance on the
screen of oscilloscope changes depending on the laser power and the speed and direction of the laser
frequency scan. This effect is produced by heating the mode volume with the light power
absorbed in the material due to nonzero optical absorption. The process can be described
with two time constants, one of which is responsible for flow of heat from the mode volume to the
rest of the resonator, and the other -- for heat exchange between the resonator and external
environment. The laser scan should be fast compared with the relaxation constants and the light power
must be small to reduce the effect.

In the simplest approximation the evolution of the system can be described with a set of two
equations \cite{fomin05josab}
\begin{eqnarray} \nonumber
&& \dot E + \left [ \gamma  +i \left ( \omega -\omega_0 + \delta  \right
) \right ]E = F(t), \\
\nonumber && \dot \delta + \Gamma \delta = \Gamma \gamma \xi |E|^2,
\end{eqnarray}
where $E$ is the complex amplitude of the field in the resonator mode normalized such that $|E|^2$ is
the total energy accumulated in the mode, $\gamma$ is the optical mode half width at the half
maximum, $\omega$ is the frequency of the optical pump, $\omega_0$ is the unperturbed resonance
eigenfrequency, $\delta=\omega_0 \beta \Delta T$ is the thermal frequency shift,
$\beta=(1/n_0)(\partial n_0/\partial T)$, $\Delta T$ is the temperature deviation, $F(t)$ stands for
the external optical pump, $\Gamma=2D/h^2$ characterizes thermal relaxation rate, $D= \lambda^*/(\rho
C)$ is thermal diffusivity, $\lambda^*$ is thermal conductivity, $\rho$ is material density, $C$ is a
specific heat capacity of diamond, $h$ is the half-thickness of the mode in the direction of the
largest field gradient, which is the radial direction for the spherical resonator; and
\begin{equation} \label{xi}
\xi = \frac{\omega_0\beta n_0^2h^2}{8 \pi \lambda^* V_{eff}}
\frac{Q}{Q_{abs}}
\end{equation}
is the thermal nonlinearity coefficient, where $Q_{abs}$ is the quality factor determined by the
material absorption. The thermal nonlinearity becomes important when the power accumulated in the
resonator exceeds $\xi^{-1}$. It is easy to see from Eq.~(\ref{xi}) that thermal nonlinearity of a
diamond resonator is very small compared with the thermal nonlinearity of other optical
materials, since the thermal conductivity in diamond is significantly larger compared with thermal
conductivity of any other optical crystal.

To confirm the conclusion we performed a comparative measurement of thermal nonlinearity of two
identical resonators. One of the resonators was the diamond one, and the other was made with MgF$_2$. The
MgF$_2$ resonator was fabricated so that its $Q_{abs}$ was comparable with the Q of the diamond
resonator. In this case $\xi_{MgF2}/\xi_{diamond} \simeq 360$. It means that thermal nonlinearity
becomes visible in magnesium fluoride for pump power hundred times smaller compared with the
identical diamond resonator.

The thermal nonlinearity can be significantly suppressed if the resonators of
high transparency material are overloaded in such a way that the loaded Q becomes much smaller than
the absorption-related $Q_{abs}$ ($Q_{abs} \gg Q$).  If one is able to maintain extremely low levels
of absorption in a resonator with transparent material with $Q_{abs}$ much larger than the one of
diamond, as in the case of CaF$_2$ \cite{savchenkov07oe} and MgF$_2$, the benefit of diamond as
extreme thermal conductivity material could become irrelevant. This, however, is a challenging engineering
task, since not only the resonator material must have ultra-low loss, but  absorptive
contaminations must also be avoided during integration of the resonator.

The diamond resonators have an important advantage compared with resonators of other materials if the
goal is to create an ultra-stable cavity. The ultimate stability of monolithic resonators is limited
by fundamental thermodynamic fluctuations existing even if the resonator is kept at constant
temperature. For instance, thermodynamic fluctuation of temperature in the mode volume results in
fluctuations of the index of refraction in the mode channel that can be written as
\cite{matsko07josab}
\begin{eqnarray}
\frac{\Delta n}{n}\biggr \vert_{(\Delta T)_m} = \beta (\Delta
T)_m. \label{dn}
\end{eqnarray}
The mean square value of the temperature fluctuation is
\begin{eqnarray}
\langle (\Delta T)_m^2 \rangle = \frac{k_BT^2}{C \rho V_{eff}},
\label{dt}
\end{eqnarray}
where $k_B$ is the Boltzmann's constant, $T$ is the absolute temperature.

The spectral power  density of the fluctuation of the averaged
temperature of the mode of a toroidal resonator ($\bar T(t)$) is
\cite{matsko07josab}
\begin{eqnarray}
\label{sdt} S_{\bar T}(\Omega) \approx \frac{ k_B T^2 a^2}{12 \lambda^*
V_{eff}}
 \left [ 1+\left (\frac{a^2}{D} \frac{|\Omega|}{9\sqrt 3} \right )^{3/2}
\right ]^{-1}.
\end{eqnarray}
The spectral density is inversely proportional to the thermal conductivity of the resonator host
material. Therefore, the thermorefractive noise will be significantly smaller in a diamond resonator.
The thermal fluctuation does not depend on the loss of the resonator host material. It shows that the
existing resonator already has certain advantages over other, higher-Q, crystalline resonators.

\section{Crystalline Quartz Whispering Gallery Mode Resonator}

By adapting the mechanical polishing technique we have developed, we fabricated quartz WGM resonators \cite{ilchenko08ol} out of cylindrical crystal quartz preforms purchased from two different vendors. The mechanical polishing was realized with a diamond slurry. The polishing quality is high enough to ensure negligible surface scattering of light traveling in the WGMs. Extremity of the resonators is shaped into toroidal geometry.

\begin{figure}[htb]
\centerline{
\includegraphics[width=14.5cm]{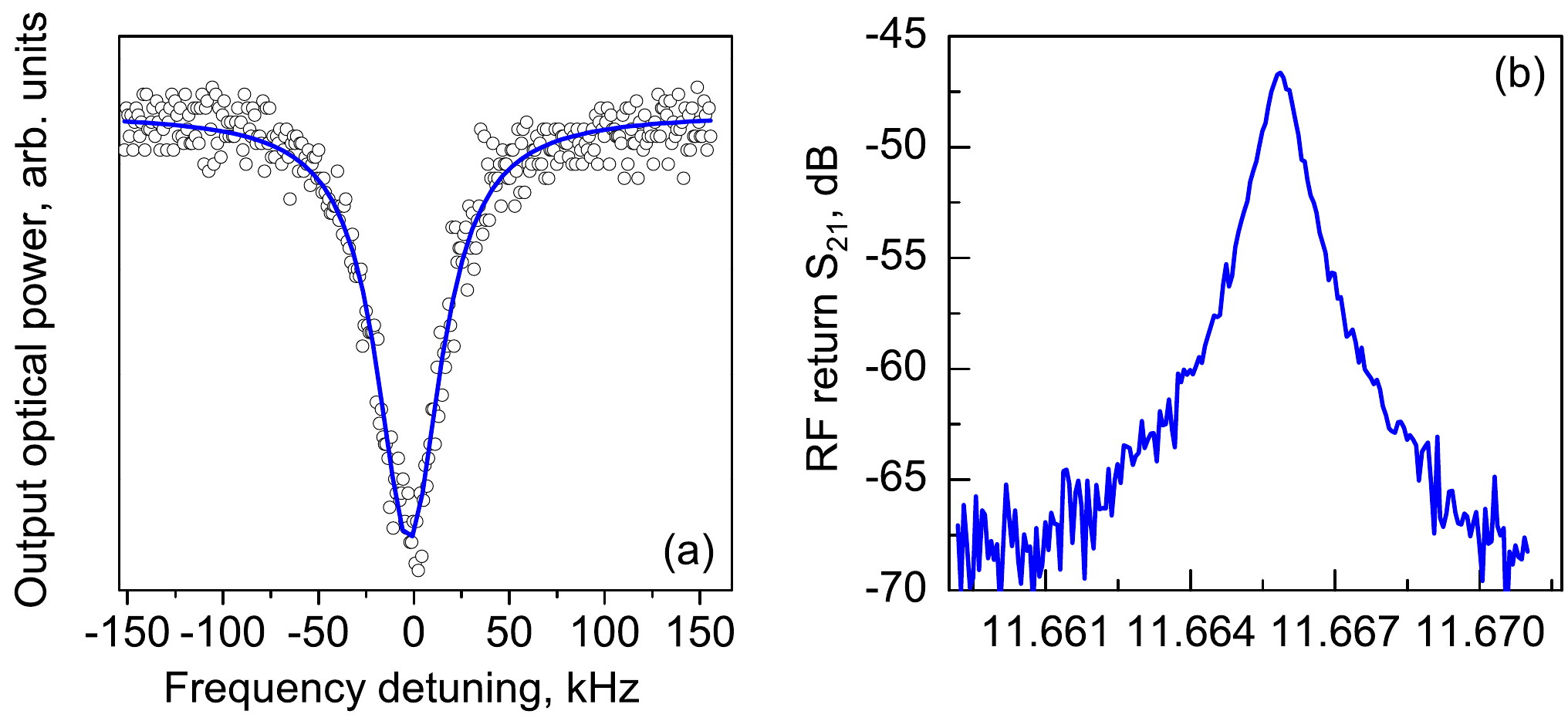}}
\caption{\label{fig1quartz} (a) A WGM resonance belonging to a z-cut quartz WGM resonator. Solid line is
a Lorentzian fit to the experimental data. The optical power sent to the mode is
small enough ($P_{in}=3 \ \mu$W) to avoid nonlinear processes. The linewidth corresponds
to $Q=5 \times 10^9$. (b)  RF return for the modulator based on the X-cut quartz resonator versus RF modulation
frequency (the ratio of the powers of the demodulated signal and the input signal). The input microwave as
well as the optical power at the high speed photodetector is equal to 1 mW.
The RF return frequency bandwidth is $\sim 0.8$~MHz at $-3$~dB. The loaded quality factor of the
optically pumped WGM is $Q=2 \times 10^8$ (optical bandwidth is 1~MHz). The optical mode contrast is 50\%.
The laser is manually tuned to the mode. The maximum RF return corresponds to $S_{21}\simeq
2 \times 10^{-5}$.}
\end{figure}

Modes in the resonators were excited using a SF11 prism coupler and a 1550~nm tunable fiber laser having 4~kHz linewidth. The modes with quality factors approaching $Q=5\times10^9$  (see Fig.~\ref{fig1quartz}a) were detected in z-cut resonators (the symmetry axis of the resonators coincides with the z-axis of quartz).

Because of the crystalline properties a Z-cut quartz resonator is not tunable by an electric field applied along its symmetry axis. We  placed metal electrodes on the top and bottom surfaces of a $100\ \mu$m thick resonator and applied a DC voltage (as done in \cite{savchenkov03el}). No measurable voltage-dependent mode shift was observed for the voltage amplitude exceeding 200~V.

To observe the electric tunability of the WGM spectrum we fabricated resonators out of x-cut quartz. The optical quality of the x-cut crystal quartz samples in our disposal was not good enough so the maximum Q-factor of those resonators was an order of magnitude lower than that of the z-cut resonators.

We have measured the value of the Pockels constant of quartz (first
measured by Pockels himself
\cite{pockels1894agw,cadybook,rosner67ao}) by applying a DC voltage
to the x-cut resonator's top and bottom surfaces and measuring the
frequency shift of the WGMs \cite{savchenkov03el}. For instance,
the TE mode family shifts by $\delta \nu = 78$~MHz when we apply
$V=133$~V to the $h=220$~$\mu$m thick disk, so that $r_{11}=2
\delta \nu h/(V \nu n^2)$, where $\nu=1.94\times10^{14}$~Hz is the
carrier frequency of the pump light and $n=1.53$ is the index of
refraction of the material at 1550~nm. The measured value is
$r_{11}=0.57$~pm$/$V. Interestingly, not only TE, but also TM modes experience frequency shifts when the electric field is applied. This basically means that the modes in our resonator do not
have the distinct parallel or orthogonal direction of the electric
field with respect to the symmetry axis of the resonator.

Electrical tunability of quartz WGM resonators enables WGM
electro-optical modulators \cite{cohen01sse-b,ilchenko03josab}
(feasibility of quartz electro-optical modulators have been noted
previously \cite{eden63ao,kaminow66ao,pursey68bjpd}). We have
fabricated such a modulator using a standard technique utilized
previously with lithium-niobate WGM modulators
\cite{ilchenko03josab}. We mounted a strip line RF resonator
on top of the  resonator sitting on a polished brass
electrode, and tuned the RF resonance frequency such that it became
identical to the free spectral range (FSR) of the optical
resonator. The strip line was placed at the resonator
circumference to ensure an overlap between the RF field and the
light confined in the modes. The resultant RF strip line resonator
covered approximately 190$^o$ of the WGM resonator perimeter. The
full width at the half maximum (FWHM) of the RF mode was 215~MHz,
the RF resonant frequency was $11.7$~GHz, hence the RF quality
factor was $Q_{RF}\simeq 54$.

To operate the modulator we  sent laser radiation to a WGM and
a continuous wave RF signal to the strip line resonator. The nonzero
$r_{11}$ of quartz ensures the wave mixing process between
light and the RF signal. The wave mixing results in generation of
optical sidebands resonant with the optical modes separated from
the pumped WGM by an integer number of FSRs. The first order
sidebands grow first, followed by the higher order sidebands.
We demodulated the light escaping the resonator by means of a
fast photodiode and observed the RF signal using an RF spectrum
analyzer (Fig.~\ref{fig1quartz}b). It is possible to find the parameters
of the modulator using this signal.

The RF return of the modulator can be expressed as
\cite{ilchenko03josab}
\begin{eqnarray} \label{s21}
S_{21}=\xi \frac{\rho_{pd} {\cal R}^2 P_{in}^2}{P_{sat}} , \\
P_{sat}=\frac{n_{RF}^2
\omega_{RF}{\cal V}_{RF}} {8 \pi Q^2 Q_{RF} r_{11}^2n^4 \eta^2},
\end{eqnarray}
where $1>\xi>0$ is the numerical parameter taking into account the
coupling efficiency of the light into the resonator as well as the
type of the modulation (an all-resonant modulator produces mostly
phase modulated light zeroing down $S_{21}$), $P_{sat}$ is the RF
saturation power of the modulator, $\rho_{pd}$ is the resistivity
of the photodiode, ${\cal R}$ is the responsivity of the
photodiode, $P_{in}$ is the power of the light, $Q$ and $Q_{RF}$
are the quality factors of the optical and microwave modes,
$r_{11}$ is the electro-optical coefficient of the resonator host
material, $n$ and $n_{RF}$ are the indices of refraction of the
material, ${\cal V}_{RF}$ is the volume of the microwave field,
$\eta = (1/V_e)\int |\Psi_e|^2\Psi_{RF} dV <1$ is the overlap
integral of the optical and microwave fields, $|\Psi_e|^2$ and
$|\Psi_{RF}|^2$ are the spatial distributions of the power of the
optical and microwave fields respectively, $(1/V_e)\int |\Psi_e|^2
dV = 1$, and $\omega_{RF}$ if the RF frequency.

To find the saturation power we use $Q=2 \times 10^8$,
$Q_{RF} =54$, $\omega_M = 2\pi \times 11$~GHz, $\eta=0.5$,
$r_{11}=0.57$~pm$/$V=$1.3\times10^{-9}$~esu, $n_e=1.53$,
$n_{RF}=2.15$, ${\cal V}_{RF} = 2 \times 10^{-4}$~cm$^3$ (the
volume is given by the 100~$\mu$ wide strip line resonator having
$\pi R$ length and 220~$\mu$m thickness). Then we estimate $P_{sat}
\simeq 51$~mW. We also know from the experiment that $S_{21}\simeq
2 \times 10^{-5}$, $P_{in}=1$~mW, $\rho_{pd}=50$~Ohm, and ${\cal
R}=0.8$~A$/$W. Using these values and the calculated value for the
saturation pump power we obtain $\xi=4 \times 10^{-2}$. Because
$\xi$ is small our modulator generate mostly phase modulated light.

The electro-optical modulator based on the quartz WGM resonator
will become extremely important if the quality factor of the optical resonator
fabricated from the x-cut material can be
significantly increased. This will cause the saturation power of such a
resonator to be low, so the modulation will be very efficient.
The modulator will then also have an extremely narrow bandwidth. Such properties
are important in harmonically mode locked lasers
\cite{harvey93ol,gee05ptl} and coupled opto-electronic oscillators
\cite{yao97ol}. Indeed, optical etalons have been inserted into the
optical loop to improve the performance of harmonically actively
mode-locked lasers \cite{harvey93ol,gee05ptl}, and electro-optical
modulators are the key parts of those lasers. The quartz WGM
modulator is able to substitute both the modulator and the etalon
in the mode-locked lasers. The low saturation power of quartz
modulator will eliminate the requirement of
high power RF amplifiers in oscillators, while the narrow filter function will reduce
the phase noise.

A laser locked to a narrow resonance high-finesse resonator is the commonly used  source of stable narrow-linewidth optical signals.  Extremely narrow linewidth of the WGMs in quartz resonators will
allow realizing lasers with high short-term stability, while
electro-optical tunability of the WGM will allow tuning or
modulating the frequency of the lasers.

Finally, let us discuss the nonlinear optical properties of the
quartz resonators. It is known that quartz has several Raman lines
\cite{tannenwald66s,scott67pr,briggs77prb} and possesses a large cubic
nonlinearity ($n_2=1.7 \times 10^{-14}$~esu, $1.4$ times larger
than  fused silica \cite{maker65pr,meredith81prb}).
Therefore, we should expect to observe
stimulated Raman scattering (SRS) as well as four-wave mixing
in the quartz resonators as previously observed in fused silica resonators. To examine this, we amplified our laser
and pumped the quartz resonators with $40$~mW light. The only
effect observed was the SRS. No hyper-parametric
oscillations based on  four-wave mixing were detected. The reason
for the suppression of the effects will be the subject of further investigations.

\section{Conclusion}
We have presented results concerning  fabrication of high-Q sapphire, diamond, and crystal quartz
whispering gallery mode resonators, and demonstrated the highest reported Q-factor in a monolithic
resonators fabricated out of the materials. The resonators can be useful in a variety of
scientific and technical applications, including fabrication of ultra-stable optical etalons,
cavity-stabilized lasers, and opto-electronic radio-frequency oscillators.

\section*{Acknowledgments}
The authors acknowledge support from Defense Sciences Office of Defense Advanced Research Projects Agency under contract No. W911QX-12-C-0067 as well as support from Air Force Office of Scientific Research under contract No. FA9550-12-C-0068.



\end{document}